# AI-Assisted Thin Section Image Processing for Pore-Throat Characterization in Tight Clastic Rocks

## Muhammad Risha[1*]

[1]Department of Marine, Earth, and Atmospheric Sciences, North Carolina State University, Raleigh, NC 27695, USA

## ABSTRACT

The characterization of pore-throat structures in tight sandstones is crucial for understanding fluid flow in hydrocarbon reservoirs and groundwater systems. Both thin-section and Mercury Intrusion Capillary Pressure (MICP) offer insights rock petrophysical parameters. However, thin-section analysis is limited by its 2D nature and subjective interpretation, while MICP provides 3D pore-throat distributions, it lacks direct visualization of pore morphology. This study evaluates AI-assisted thin-section image analysis for pore-throat characterization by comparing its results to MICP-derived measurements. A machine learning-based workflow was developed using color thresholding, K-Means clustering, and medial axis transformation to segment pore structures in thin-section images. Throat width, porosity, and permeability were quantitatively assessed against MICP to determine the accuracy and reliability of the technique. The analysis of 26 sandstone samples outlined differences between the two methods. Thin-section analysis showed porosity values from 1.37% to 53.37%, with average pore-throat sizes between 5.63 µm and 30.09 µm, while permeability estimates ranged from 0.01 mD to 344.35 mD. Correlation analysis showed moderate agreement for throat size (r=0.62) and permeability (r=0.61), but weaker for porosity (r=0.32), highlighting the differences in how each method captures pore connectivity. Results demonstrate that the AI-assisted segmentation provides a scalable and reproducible approach but is constrained by thin-section imaging resolution. While MICP remains reliable for permeability evaluation, its comparison with AI-driven image analysis helps assess the reliability of the method. Future research should refine segmentation algorithms, incorporate pretrained data to validate AI-derived pore-throat attributes for improved reservoir quality assessment and predictive modeling.

## 1. INTRODUCTION

The pore structure of sandstones is a critical factor influencing fluid storage and transport properties in subsurface reservoirs. The size, shape, and connectivity of pores and throats control key parameters such as permeability, capillary pressure, and fluid retention, which directly impact hydrocarbon production, groundwater movement, and carbon sequestration (Izadpanahi et al., 2024; Jerauld & Salter, 1990; Kim & Lee, 2018; Lai et al., 2018). Understanding these properties requires accurate pore-throat characterization, traditionally performed through thin-section petrography and laboratory petrophysical measurements (Cannon, 2015; Stadtmüller & Jarzyna, 2023). While petrographic analysis provides valuable insights into rock fabric and pore morphology, its reliance on manual interpretation makes it time-consuming and prone to subjectivity, potentially leading to inconsistencies in large datasets (Li et al., 2019; Vasquez et al., 2018).

Recent advances in digital image processing have allowed for more objective and automated characterization of pore networks. Computational methods, particularly machine learning-based image segmentation, have demonstrated significant potential in analyzing high-resolution thin-section images, enabling faster and more reproducible pore-throat quantification (Berrezueta et al., 2015; Han & Liu, 2024; Thomson et al., 2018). Unsupervised learning approaches such as K-Means clustering have been widely applied in geological image processing, offering a practical alternative to deep learning methods that require extensive labeled datasets. In addition, morphological operations and watershed segmentation have proven effective in refining pore boundary delineation and pore-throat extraction, helping to mitigate illumination-related artifacts and enhance segmentation accuracy. These methodologies allow for a detailed and scalable representation of pore-throat systems, providing a robust framework for integrating image-based analysis with traditional petrophysical measurements (Berrezueta & Kovacs, 2017; Jiu et al., 2021).

Despite the advantages of AI-assisted segmentation, its effectiveness in quantifying pore-throat structures must be validated against established petrophysical techniques. MICP analysis is widely used for pore-throat size distribution measurements and remains a benchmark for evaluating permeability trends in porous media (Greene et al., 2022). A direct comparison between AI-extracted pore-throat attributes and MICP-derived parameters provides a means to evaluate the reliability of computational image analysis for reservoir characterization. This study presents an AI-assisted workflow for pore-throat characterization in sandstones using thin-section image analysis. The methodology employs a combination of color-based thresholding in the HSV color space, K-Means clustering for segmentation, and medial axis transformation for pore-throat width estimation. The extracted pore and throat attributes are quantitatively compared with MICP measurements to assess the accuracy and applicability of this approach. By integrating machine learning-based segmentation with traditional petrophysical analysis, this research aims to establish an efficient, reproducible method for pore-throat characterization, providing a systematic alternative to manual petrographic interpretation while maintaining compatibility with established petrophysical techniques (Khan & Khanal, 2024).

## 2. METHODOLOGY

### 2.1 Thin-Section Image Acquisition

Sandstone samples were collected from outcrops on Labuan Island, representing three distinct formations: Belait, Temburong, and Crocker. These formations are key components of the regional stratigraphy and encompass the primary tight sandstone facies in the area, making them suitable for studying pore-throat characteristics in various depositional environments (Madon, 1997; Risha & Douraghi, 2021). Standard petrographic techniques were used to prepare thin sections, ensuring uniform thickness and optimal optical clarity. To enhance the visibility of pore spaces and preserve structural integrity during sectioning, the samples were impregnated with blue-dyed epoxy. High-resolution optical microscopy was employed to capture images under plane-polarized and cross-polarized light, enabling detailed examination of mineralogical composition and pore structure. Digital images were acquired at consistent magnifications to maintain uniform spatial resolution across all samples (Hughes & Cuthbert, 2000; Risha et al., 2023). Calibration was performed using stage micrometry to establish precise pixel-to-micron conversion factors based on the magnification-scale conversion, ensuring accurate pore spaces measurements.

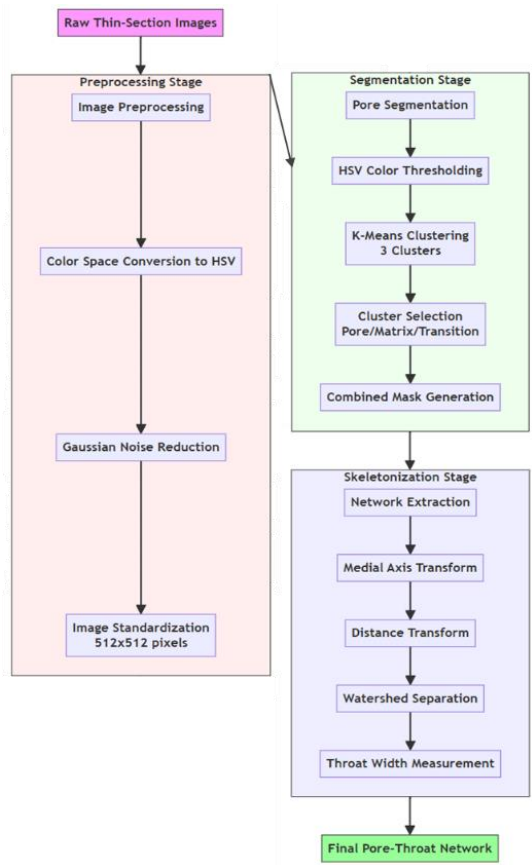

Figure 1. AI-assisted image processing workflow for pore-throat characterization. The workflow consists of three primary stages

## 2.2 AI-Based Image Processing Workflow

To automate pore-throat characterization, an AI-assisted image processing workflow was developed using Python. The workflow integrates deep learning models and computational image analysis techniques, using libraries such as OpenCV, Scikit-Image, TensorFlow/Keras, and SciPy for efficient image processing and segmentation. The workflow consists of three main stages: image preprocessing, segmentation, and pore-throat quantification.(Bukharev et al., 2018; Suo et al., 2024) (Figure 1).

## 2.3 Image Preprocessing

Thin-section images often contain noise, uneven illumination, and background artifacts that can affect segmentation accuracy. To enhance image clarity, color-based thresholding was applied in the Hue-Saturation-Value (HSV) color space, specifically targeting blue-dyed epoxy-filled pore spaces. Gaussian filtering was used to reduce noise while preserving fine pore structures (Jiu et al., 2021). The images were standardized to a resolution of 512 × 512 pixels to maintain consistency for the model processing (Feng et al., 2020).

## 2.4 Pore Segmentation Using K-Means Clustering

Segmentation was conducted using an unsupervised clustering approach via K-Means, which grouped pixels based on color similarity in the HSV space. This method was employed to refine the delineation of pore spaces and minimize the reliance on manually set threshold values. The clustering process helped differentiate pore regions from the mineral matrix, providing a more robust classification of pore spaces. A three-cluster approach was used, with one cluster representing the pore spaces, another the surrounding rock matrix, and the third accounting for transition zones. The cluster most representative of the blue-impregnated pores was selected and combined with the HSV-based threshold mask for improved segmentation fidelity (Liu & Ren, 2022). The effectiveness of K-Means clustering for color-based image segmentation has been widely demonstrated in image processing applications, including geological and biomedical imaging (Jumb et al., 2014; Sari et al., 2020). The extracted pore networks were further processed using distance transformation, watershed segmentation, and skeletonization, which have been recognized as effective techniques for pore-throat structure identification and connectivity assessment (Wang et al., 2013; M. Zhang et al., 2021).

## 2.5 Pore and Throat Skeletonization

Once pores were segmented, the pore-throat network was extracted using skeletonization techniques. The medial axis transform was applied to generate a one-pixel-wide representation of the interconnected pore network, allowing for the identification of pore throats. The distance transform was used to calculate throat widths by measuring distances from the skeletonized structure to the nearest boundary of segmented pores. The watershed algorithm was then used to separate closely spaced pore structures, ensuring accurate throat width measurement (Y. Zhang et al., 2012).

To further illustrate the effectiveness of the segmentation and skeletonization process, Figure 2 presents a zoomed-in view of a sandstone thin-section, demonstrating the stepwise refinement of pore-throat identification. The figure highlights the transition from the original thin-section image to the progressively refined segmentation stages, culminating in the final skeletonized pore-throat network.

This visualization provides a clear representation of how HSV thresholding, K-Means clustering, and morphological operations enhance pore detection and connectivity mapping. The skeletonized structure extracted in the final step serves as the foundation for quantitative throat width measurements, ensuring accurate pore-throat characterization for further permeability analysis.

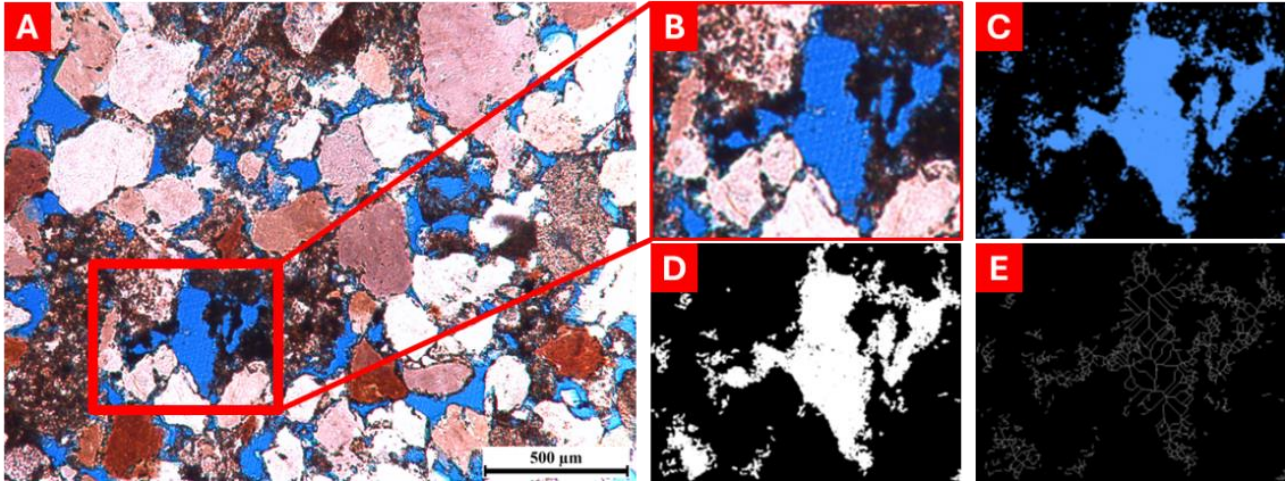

Figure 2. Stepwise pore segmentation improvement in a zoomed-in section of a sandstone thin-section image. (A): Original thin-section image, with a red box highlighting the magnified area used for segmentation analysis. (B): Zoomed-in view of the selected area, showing blue-dyed epoxy-filled pore spaces within the mineral matrix. (C): HSV-based color thresholding applied to isolate pore structures, providing an initial classification. (D): Binary segmentation output after refining the thresholded image, enhancing pore-matrix separation. (E): Final skeletonized pore-throat network, extracted using medial axis transformation, enabling quantitative pore-throat analysis.

### 2.6 Pore-Throat Measurement and Quantification

Thin-section image analysis provides a valuable method for quantifying pore structures, but it is inherently limited by optical resolution. The resolution of petrographic thin-section imaging typically restricts the detection of the smallest pore features, as finer details can be lost due to diffraction limits and pixel resolution constraints. While high-resolution imaging techniques such as scanning electron microscopy (SEM) and X-ray microtomography (μCT) provide better pore visualization, thin-section microscopy is constrained in its ability to resolve pores smaller than a certain threshold (Alves et al., 2014). To ensure reliable pore-throat quantification, this study excludes pores smaller than 10 μm from the analysis. This threshold was chosen based on the optical limitations of thin-section imaging and the need to avoid overestimating porosity due to artifacts and noise introduced at sub-resolution scales. Studies have shown that pore-throat measurements below this range in 2D thin-section images tend to be highly uncertain, as the boundaries of such small features may be inaccurately segmented due to pixelation and blurring (Dong et al., 2019). Additionally, excluding pores below this size range helps minimize misclassification of micro-pores and fine mineral matrix textures, which can otherwise be incorrectly identified as part of the connected pore network (White et al., 1998).

The measured parameters include pore size distribution, throat width, aspect ratio, and circularity. Porosity ($\phi$) is determined as the ratio of total pore area to the total rock area in the thin-section image:

$$\phi = \frac{A_p}{A_t} \tag{1}$$

where $A_p$ is the total pore area and $A_t$ is the total image area. Pore size is estimated using the Feret diameter $D_f$, which represents the longest distance between any two points within a pore:

$$D_f = max(d_{ij}) \tag{2}$$

where $d_{ij}$ is the Euclidean distance between two boundary points of the pore. Throat width is determined using the Euclidean Distance Transform (EDT) applied to the skeletonized pore network. The local throat width at each point along the skeleton is given by:

$$w_t = 2 \times min\big(d(x,B)\big), \ \forall x \in S \tag{3}$$

where d(x,B) is the distance from a skeleton point x to the nearest boundary B, and S represents the skeleton of the pore-throat network. Aspect ratio AR provides insights into pore elongation and potential anisotropy in the reservoir matrix. It is defined as:

$$AR = \frac{L_{max}}{L_{min}} \tag{4}$$

where $L_{max}$ and $L_{min}$ are the major and minor axes of the fitted ellipse to the pore shape. Circularity, which quantifies how close the pore shape is to a perfect circle, is computed based on (Ekneligoda & Zimmerman, 2008):

$$C = \frac{4\pi A}{P^2} \tag{5}$$

where A is the pore area and P is the perimeter. A value of 1.0 corresponds to a perfect circle, while values approaching zero indicate highly irregular pore shapes. Permeability estimation was performed using the Kozeny-Carman equation (Chapuis & Aubertin, 2003):

$$K = \frac{\phi^3}{S^2(1-\phi)^2} \tag{6}$$

where k is permeability, $\phi$ is porosity, and S is the specific surface area. The pore-throat coordination number, A higher coordination number suggests a more interconnected pore system, which correlates with higher permeability and fluid flow efficiency and (Z) is estimated based on (Xu et al., 2022):

$$Z = \frac{2E}{V} \tag{7}$$

where E is the number of pore throats (edges) and V is the number of pores (vertices) in the extracted pore network graph. By incorporating these equations, this study ensures a robust and reproducible pore-throat characterization methodology, minimizing uncertainties associated with resolution limitations while maximizing the accuracy of permeability predictions (Beckingham et al., 2013).

## 2.7 Validation with MICP

MICP is a well-established technique for quantifying pore-throat size distributions and remains an industry benchmark for permeability estimation (Gao et al., 2019). Statistical correlation analyses were performed to compare AI-derived pore-throat parameters with MICP-derived distributions, ensuring the reliability and applicability of the deep learning approach in sandstone reservoir characterization (Nie et al., 2021).

To validate the accuracy of image analysis results, they were compared with MICP analysis results conducted using the Thermo Scientific AutoPore IV 9500. This instrument provides high-resolution pore-throat size distributions by applying stepwise increases in mercury intrusion pressure and measuring the corresponding volume of mercury entering the pore network. The analysis followed ISO 15901-1 standards, ensuring reproducibility and reliability in characterizing pore structures in sandstone samples.

The raw data obtained from the MICP report included applied pressure values in megapascals (MPa), cumulative mercury intrusion volumes in cubic millimeters per gram (mm³/g), and pore-throat diameters in nanometers (nm), derived from the applied pressure using the Washburn equation. Pore-throat diameters were determined using the Washburn equation (Cai et al., 2021):

$$d = \frac{-4\gamma \cos\theta}{P} \tag{8}$$

where d represents the pore-throat diameter, $\gamma$ is the mercury surface tension (0.48 N/m), $\theta$ is the contact angle, and P is the applied pressure in megapascals. Accessible porosity was calculated from the total intruded volume, corrected for compressibility effects.

Permeability was obtained from the MICP report based on a cylindrical pore model and expressed in millidarcies (mD) using the conversion below:

$$K(mD) = K(\mu m^2) \times 1013.25 \tag{9}$$

## 3. RESULTS

The pore-throat characteristics of 26 sandstone samples were analyzed using MICP measurements and thin-section image processing techniques. These methods provided complementary datasets, allowing for a comparative assessment of porosity, average pore-throat size, and permeability across different sandstone formations. Thin-section analysis enabled a 2D pore structure evaluation, while MICP provided a 3D-based pore-throat size distribution through capillary pressure measurements.

### 3.1 Pore Structure and Segmentation Results

The thin-section image analysis reveals a wide variation in pore structures among the sandstone samples studied. Thin-section images were processed using an AI-assisted segmentation workflow to extract pore and throat structures. The process involved three major steps: (1) acquiring high-resolution thin-section images, (2) performing segmentation using a watershed-based algorithm to isolate pore spaces, and (3) applying skeletonization to trace the pore-throat network. The image processing workflow illustrates the stepwise transformation of a thin-section image into a segmented pore network and a skeletonized pore-throat structure (Figure 3).

The original thin-section image highlights pore spaces with blue-dyed epoxy, distinguishing them from the surrounding mineral matrix. Through segmentation, the pore network is extracted, where blue regions correspond to identified pore spaces. Further processing using medial axis transformation refines the pore-throat structure into a one-pixel-wide skeleton, enabling quantification of throat widths, connectivity, and overall pore structure. Thin-section analysis on the 26 samples showed porosity values ranging from 1.37% to 53.37%, with an average pore-throat size between 5.63 μm and 30.09 μm. Permeability estimates obtained from thin-section images varied significantly, spanning from 0.01 mD to 344.35 mD. The segmentation process provided detailed insights into pore connectivity, highlighting differences in pore-throat density and arrangement across samples.

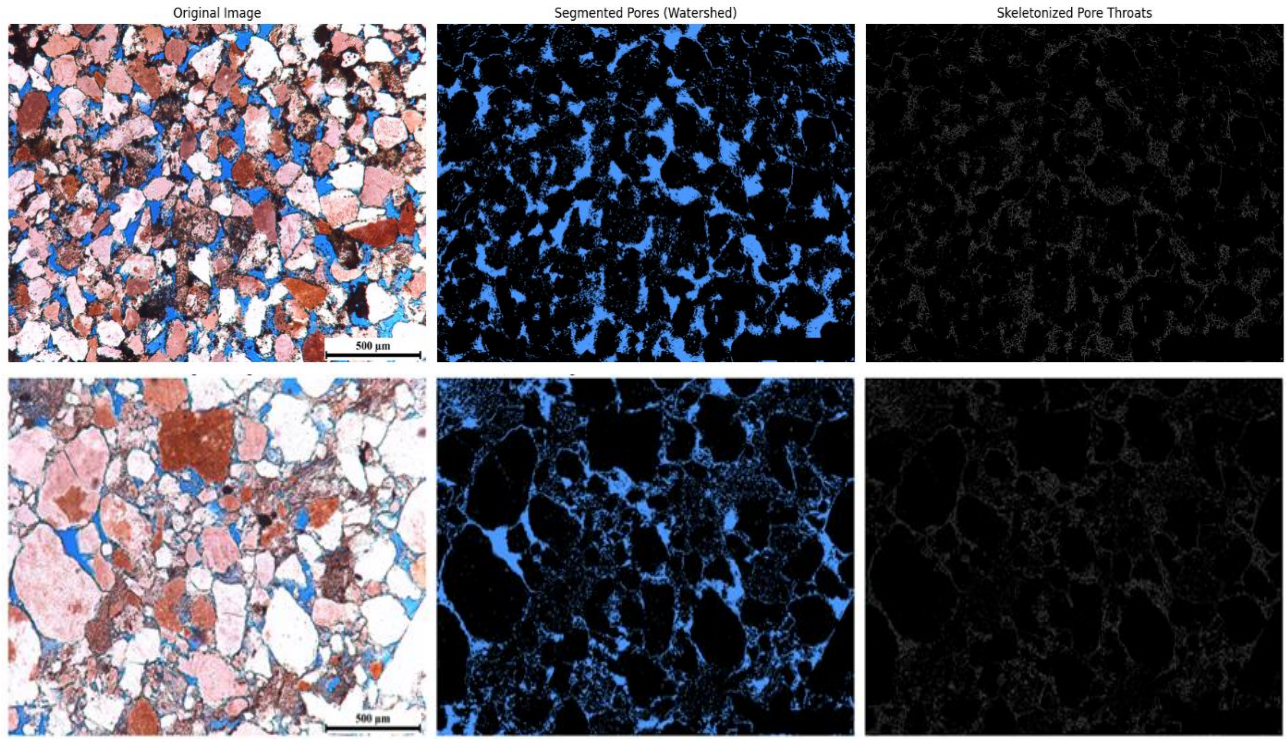

Figure 3. Segmentation and skeletonization results for two examples of sandstone samples. The left column shows the original thin-section images, where blue-dyed epoxy highlights pore spaces. The middle column presents segmented pores using HSV thresholding and K-Means clustering, refining pore identification. The right column displays the skeletonized pore-throat networks, obtained via medial axis transformation, allowing for quantitative throat width analysis and connectivity assessment. This comparison highlights variations in pore structure and connectivity between different sandstone samples.

## 3.2 Correlation Between MICP and Thin-Section Measurements

A direct comparison of MICP and AI-assisted thin-section image analysis was performed to evaluate the consistency of porosity, pore-throat size, and permeability measurements.

Results show variations between MICP and thin-section measurements, with some samples exhibiting significant deviations. While a moderate correlation is observed for throat size and permeability, porosity shows a weaker relationship, indicating differences in how each method captures pore connectivity and volume (Figure 4).

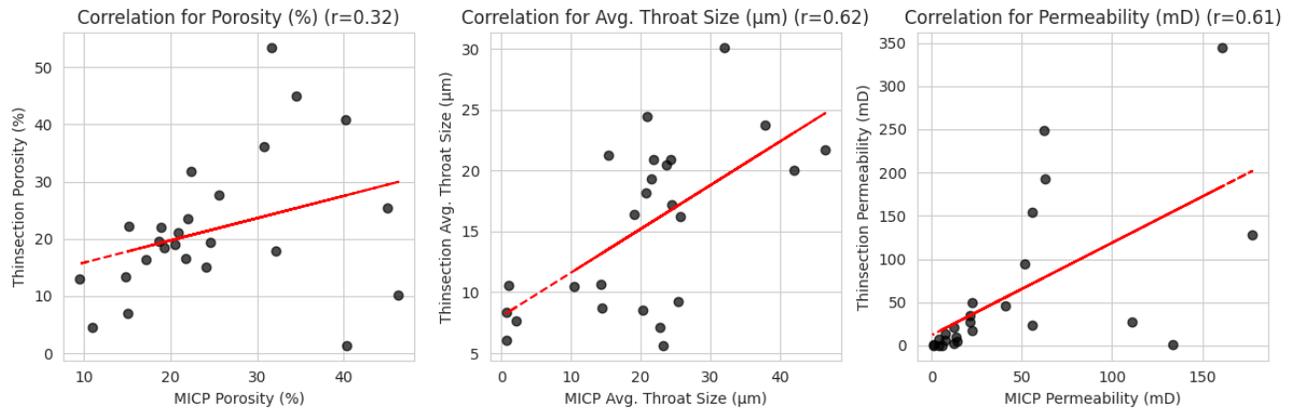

Figure 4. Correlation between MICP and thin-section image analysis for porosity, average throat size, and permeability. The scatter plots compare MICP-derived values with those obtained from AI-assisted thin-section analysis. The red dashed lines represent linear regression fits, with correlation coefficients (r) indicating the strength of the relationship.

The correlation for porosity (r=0.32) is relatively weak, suggesting significant variation between the two techniques. This discrepancy is likely due to the 2D nature of thin-section imaging, which may detect isolated pore spaces that do not contribute to the connected pore system in 3D, whereas MICP measures the total interconnected porosity within the sample. For pore-throat size, a moderate correlation (r=0.62) is observed. The trend indicates that thin-section-based measurements align with MICP, but smaller throat sizes are systematically underestimated in thin-section analysis. This limitation arises from the optical resolution constraints of thin-section imaging, which cannot capture sub-micron pore throats that are detectable in MICP. Permeability correlation (r=0.61) also shows a moderate agreement between the two methods, though some deviations occur in samples with high porosity and large throat sizes. These variations suggest that thin-section permeability estimation may overpredict connectivity, as it relies on 2D pore networks, whereas MICP accounts for 3D connectivity effects.

### 3.3 Distribution of Porosity, Throat Size, and Permeability

The statistical distribution of porosity, throat size, and permeability is presented in Figure 5, providing further insight into the differences between MICP and thin-section image analysis.

Porosity distributions indicate that MICP-derived values exhibit a broader range, while thin-section porosity tends to be slightly higher in certain samples. The box plots reveal greater variability in thin-section porosity estimates, likely due to segmentation-dependent variations and sample heterogeneity.

Pore-throat size distributions highlight the systematic underestimation of smaller throat sizes in thin-section analysis. The lower quartile of MICP measurements extends into the submicron range, whereas thin-section analysis consistently detects only larger throat sizes, reinforcing the effect of resolution limitations in 2D image-based methods.

Permeability distributions show a wider spread in thin-section-based estimates, reflecting greater uncertainty due to assumptions regarding connectivity in 2D analysis. MICP permeability remains within a more constrained range, as it measures true flow pathways rather than relying on assumptions about connectivity from isolated pore structures.

The combined findings illustrate the strengths and limitations of both methods. While thin-section analysis provides detailed pore structure characterization, its resolution constraints and 2D nature limit its ability to fully represent microporosity and connectivity. Conversely, MICP captures a more comprehensive 3D pore network, though it does not provide direct visualization of pore geometry. These factors must be considered when interpreting reservoir quality from either method.

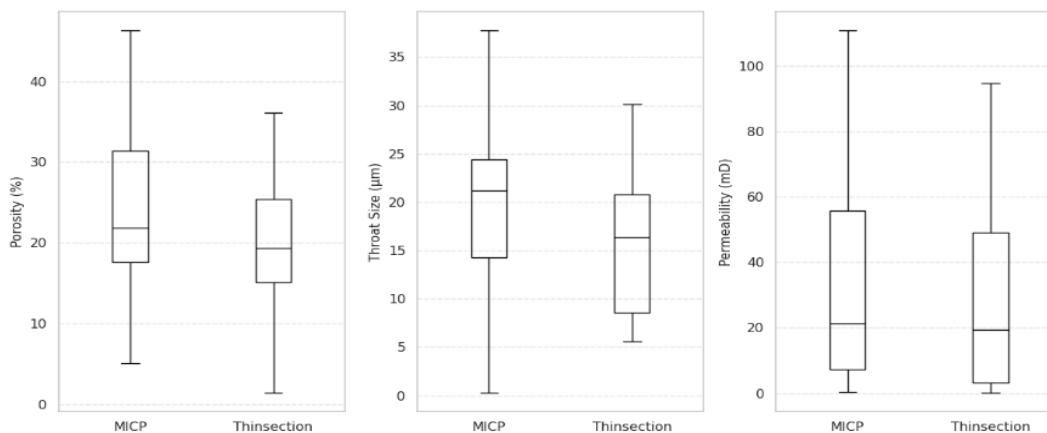

Figure 5. Comparison of porosity, throat size, and permeability distributions between MICP and thin-section analysis. Box plots illustrate the statistical distribution of porosity (left), average throat size (middle), and permeability (right) as measured using MICP and AI-assisted thin-section image processing.

## 4. DISCUSSION

The results obtained from MICP and thin-section image analysis provide insights into the pore structure, throat size distribution, and permeability variations across the studied sandstone samples. This section discusses the observed differences between the two methods, potential sources of discrepancies, and their implications for reservoir characterization.

### 4.1 Comparison of MICP and Thin-Section Image Analysis

The relationships between MICP and thin-section-derived parameters are presented in a correlation matrix of porosity, average throat size, and permeability (Figure 6). The matrix highlights the extent to which these properties are interdependent within each method and between the two approaches. The correlation between MICP porosity and thin-section porosity (r = 0.32) is relatively weak, indicating differences in how the two techniques capture pore volume. Thin-section analysis, being a 2D method, may overestimate porosity by detecting isolated pore regions that are not connected in 3D space, whereas MICP measures the total interconnected porosity. Pore-throat size exhibits a stronger correlation between methods (r = 0.62), suggesting that both approaches capture similar trends in pore structure, although thin-section analysis tends to under-represent smaller throat sizes due to optical resolution limitations. Within each method, throat size correlates well with permeability, particularly in thin-section analysis (r = 0.65). This suggests that pore-throat width plays a major role in permeability estimation when using 2D segmentation. The correlation between MICP permeability and thin-section permeability (r = 0.61) reflects moderate agreement between the two approaches. However, thin-section permeability shows a stronger correlation with thin-section porosity (r = 0.70), while MICP permeability is more closely linked to MICP throat size (r = 0.58). This suggests that MICP-derived permeability is strongly influenced by the pore-throat network, whereas thin-section-based permeability estimation depends more directly on the segmented pore space.

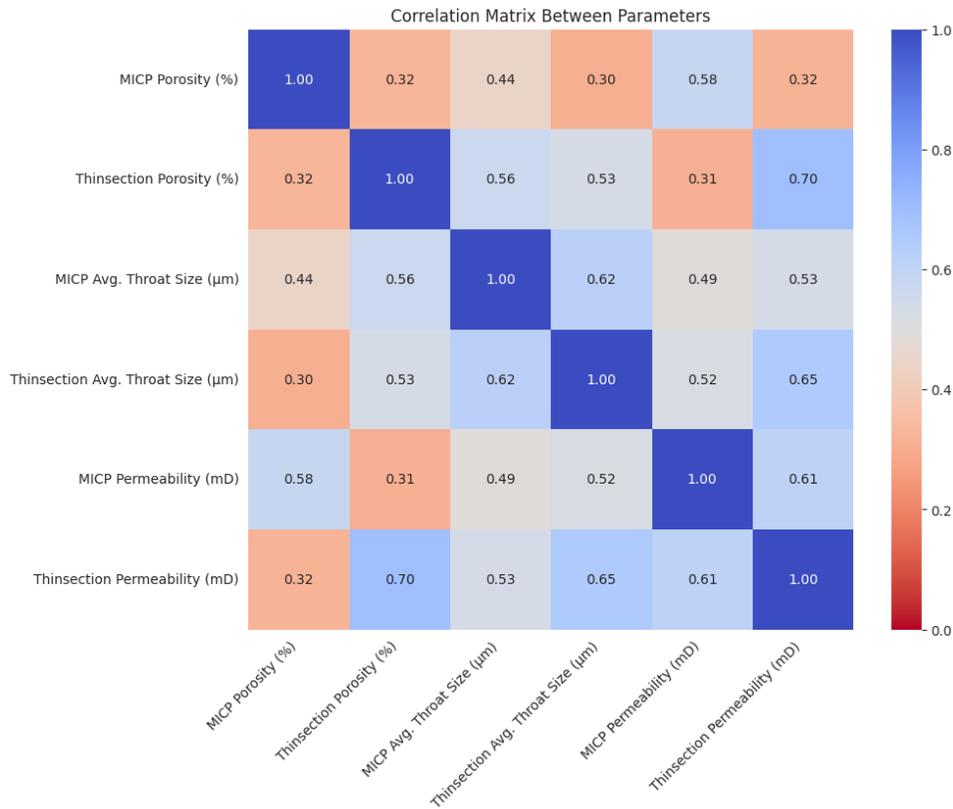

Figure 6. Correlation matrix illustrating the relationships between MICP and thin-section-derived parameters, including porosity, average throat size, and permeability. The color scale represents the strength of the correlation, with darker blue indicating stronger positive correlations and red indicating weaker or negative relationships. The matrix highlights the moderate correlation between throat size and permeability, while porosity shows weaker agreement between the two methods.

### 4.2 Variability in Pore-Throat and Permeability Estimations

A comparison between MICP and thin-section measurements reveals significant variations in porosity, throat size, and permeability across the studied samples. While both methods exhibit similar trends, individual sample deviations highlight the limitations of each approach (Figure 7). The average error between MICP and thin-section porosity is 8.53%, with the largest discrepancies occurring in highly porous samples, likely due to differences in how pore connectivity is represented in 2D versus 3D analyses. For throat size, the average error is 8.10 μm, with thin-section analysis systematically underestimating smaller throats. This discrepancy arises from the resolution constraints of thin-section imaging, which limits the detection of fine-scale pore throats that MICP can resolve through capillary pressure analysis. Permeability variations are more pronounced, with an average error of 41.86 mD. Thin-section-based permeability estimates exhibit greater variability, suggesting that 2D segmentation may overestimate connectivity in certain samples compared to MICP, which accounts for three-dimensional flow pathways.

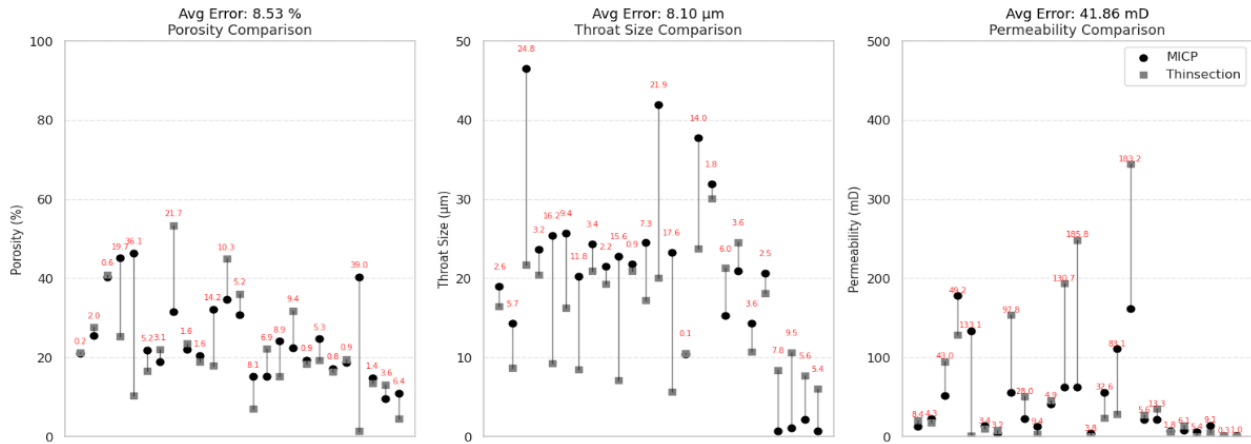

Figure 7. Comparison of porosity, throat size, and permeability measurements between MICP and thin-section analysis. Each plot displays individual sample values, with MICP measurements represented by black circles and thin-section-derived values by gray squares. The vertical lines indicate discrepancies between the two methods, with red labels showing the absolute differences.

### 4.3 Distribution Trends in Porosity, Throat Size, and Permeability

The statistical distribution of porosity, throat size, and permeability reveals key differences between MICP and thin-section measurements (Figure 8). MICP-derived porosity exhibits a broader range, whereas thin-section analysis shows greater variability, likely influenced by segmentation accuracy and heterogeneity in pore structures. For throat size, thin-section analysis primarily detects larger pores, while MICP captures a more extensive range, including sub-micron-scale features. This discrepancy highlights the resolution limitations of thin-section imaging in resolving smaller pore throats. Permeability distributions indicate that thin-section-based estimates have a wider spread, suggesting that 2D connectivity assumptions contribute to greater variability in calculated permeability values.

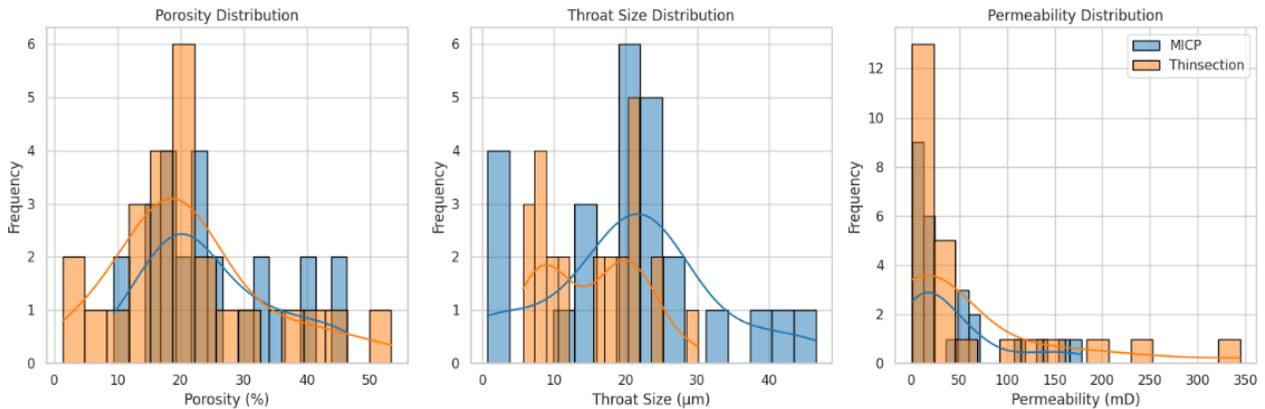

Figure 8. Histogram distributions of porosity, throat size, and permeability for MICP and thin-section analysis. The blue bars represent MICP measurements, while the orange bars correspond to thin-section-derived values. Kernel density estimation (KDE) curves overlay the histograms, illustrating the spread and central tendencies of each parameter. The porosity distribution shows that MICP exhibits a broader range, whereas thin-section analysis demonstrates higher variability. Throat size distributions reveal that thin-section analysis predominantly captures larger throats, while MICP extends into the sub-micron range. Permeability distributions indicate that thin-section estimates have a wider spread, suggesting increased variability due to 2D connectivity assumptions.

**4.4 Implications for Reservoir Characterization**

The observed differences between MICP and thin-section image analysis have implications for reservoir quality assessment. While thin-section analysis provides direct visualization of pore networks, its limitations in resolving micro-scale features and connectivity introduce uncertainties. MICP, on the other hand, offers a more comprehensive measure of permeability but lacks direct pore morphology representation. The combination of both methods enhances reservoir characterization by integrating microstructural insights from thin-section imaging with permeability estimates from MICP.

**4.5 Future Improvements and Recommendations**

To enhance the reliability of thin-section image analysis, future studies should consider integrating high-resolution imaging techniques, such as SEM or X-ray microtomography, to complement 2D segmentation. Additionally, refining segmentation algorithms and incorporating machine learning approaches could further improve the accuracy of pore-throat identification and connectivity estimation. A hybrid approach that combines thin-section imaging with MICP-derived permeability models could provide a more robust framework for reservoir quality evaluation.

**5. CONCLUSION**

This study presents a comprehensive evaluation of AI-assisted thin-section image analysis for pore-throat characterization in tight sandstones. The results demonstrate that AI-based segmentation provides a scalable and reproducible alternative to manual petrographic interpretation, offering direct visualization of pore morphology and connectivity. However, the two-dimensional nature of thin-section imaging introduces inherent limitations, particularly in resolving sub-micron-scale pore throats and accurately capturing three-dimensional connectivity.

The workflow applied on sandstone samples showed demonstrated strong potential for automated pore-throat characterization, though some systematic errors were observed. Error analysis showed that AI-based segmentation overestimated porosity by an average of 8.53%, primarily in heterogeneous samples where isolated pores were occasionally classified as interconnected voids. Pore-throat widths were underestimated by an average of 8.10 µm, likely due to the resolution constraints of thin-section imaging and challenges in delineating fine-scale throats. Permeability estimates exhibited an average error of 41.86 mD, reflecting the influence of 2D connectivity assumptions on flow path estimations.

Despite these challenges, the AI-assisted approach provided consistent and reproducible pore-throat quantification, with moderate correlation between AI-extracted throat sizes and permeability estimates (r = 0.62 and r = 0.61, respectively). These results highlight the effectiveness of AI-driven segmentation for large-scale pore analysis, while also emphasizing the need for refinements in segmentation algorithms, integration with high-resolution imaging techniques (e.g., X-ray microtomography), and multi-scale validation approaches. With further optimization, AI-based image analysis can serve as a valuable complement to traditional petrophysical methods, improving reservoir characterization and pore-throat predictions in tight sandstone formations.

This study highlights the importance of combining image analysis with traditional petrophysical techniques for a more comprehensive pore network characterization. Developing and refining pretrained models on a thin-section dataset is very important for improving similar methods. By integrating AI techniques with reservoir evaluation workflows, future advancements can enhance fluid flow predictions, improve reservoir quality assessments, and support hydrocarbon recovery strategies in tight sandstone formations.

## Author Contributions

"The author, M.R., was responsible for conceptualization, methodology, software development, validation, formal analysis, investigation, writing—original draft preparation, review and editing, visualization, and supervision. The author has read and approved the final manuscript."

## Funding

"This research was funded by the YUTP grant (015LC0-060)."

## Data Availability Statement

"The data supporting the findings of this study are available upon request from the corresponding author."

## Acknowledgments

"The author acknowledges the financial support from the Institute of Hydrocarbon Recovery, Universiti Teknologi PETRONAS."

## Competing Interests

"The author declares no relevant financial or non-financial competing interests that could have influenced the research findings."